\documentstyle[12pt]{article}
\newcommand{\beq}{\begin{equation}}
\newcommand{\eeq}{\end{equation}}
\newcommand{\beqs}{\begin{eqnarray}}
\newcommand{\eeqs}{\end{eqnarray}}
\newcommand{\tr}{{\mathrm tr\;}}
\newcommand{\Tr}{{\mathrm Tr\;}}
\newcommand{\dd}{{\mathrm d}}
\newcommand{\ee}{{\mathrm e}}
\newcommand{\fe}{f_{\mathrm even}(T)}
\newcommand{\fo}{f_{\mathrm odd}(T)}
\newcommand{\ff}{f(T)}
\newcommand{\de}{\partial}
\begin{document}
\begin{titlepage}
\begin{flushleft}
       \hfill                      {\tt hep-th/9904013}\\
       \hfill                       SISSA ref. 33/99/EP\\
\end{flushleft}
\vspace*{2mm}
\begin{center}
{\bf \LARGE Non-Critical Type 0 String Theories and}\\
{\bf \LARGE  Their Field Theory Duals  \\ }
\vspace*{8mm}
{\large Gabriele Ferretti\footnote{\tt ferretti@sissa.it}}\\
\vspace*{2mm}
{\large Jussi Kalkkinen \footnote{\tt kalkkine@sissa.it}}\\
and \\
\vspace*{2mm}
{\large Dario Martelli\footnote{\tt dmartell@sissa.it}}\\
\vspace*{4mm}
{\it SISSA, Via Beirut 2 Trieste 34014, Italy\\}
{\it and\\}
{\it INFN, Sezione di Trieste\\}
\vspace*{5mm}
\end{center}

\begin{abstract}
In this paper we continue the study of the 
non-critical type 0 string and 
its field theory duals. We begin by reviewing
some facts and conjectures 
about these theories.
We move on to our proposal for the type 0 effective
action in any dimension, its RR fields and their Chern--Simons 
couplings. We then focus on the case without 
compact dimensions and study its field theory duals. 
We show that one can parameterize all
dual physical quantities in terms of a finite number of unknown 
parameters. By making some further assumptions on the tachyon 
couplings, one can still make some ``model independent'' statements.

\end{abstract}
\vskip.5cm{\it This work is dedicated to the memory of Mauri Miettinen.}
\end{titlepage}

\section{Introduction}

Polyakov's proposal~\cite{Polyakov:a} of using type 0A/0B string theory for 
describing non-supersymmetric Yang-Mills theories is a promising and
novel attempt at formulating the QCD string and provides an alternative 
way of extending the AdS/CFT correspondence~\cite{Maldacena, Gubser:a,
Witten:a} beyond the realm of supersymmetric field theories.

Simply put, the proposal postulates using a closed string theory in $d\leq 10$
space-time dimensions with
world-sheet supersymmetry and a diagonal GSO projection that removes
all space-time fermions, thus yielding a non-supersymmetric theory in 
target space. 
One can refer to such theories for which 
$d<10$ as ``non-critical'' although the name can be confusing because Weyl 
invariance is recovered once the conformal factor of the world-sheet metric
is counted among the other space-time coordinates. Throughout this paper,
$d$ will always include such Liouville mode and hence Weyl invariance
is retained, at least at the lowest level in the sigma-model expansion.

So far, most of the literature following~\cite{Polyakov:a} has analyzed
the most conservative case of non-supersymmetric theories in 
$d=10$~\cite{Klebanov:a, Klebanov:b, Klebanov:c, Minahan:a, 
Minahan:b, Alishahiha}.
Such theories~\cite{Dixon, Seiberg} are often referred to as type 0A or 0B, 
depending on the particular GSO projection employed. Their open string 
descendants were analyzed in~\cite{Bianchi, Sagnotti, Angelantonj, Bergman}.

However, it is our opinion that the ``non-critical'' 
scenario~\cite{Alvarez:a, Ferretti:a, Alvarez:b, Costa, Armoni, Zhou} 
must be taken 
seriously and will give rise to additional interesting models that are
not accessible at $d=10$. Throughout the paper we shall continue to refer 
to these lower dimensional theories as type 0A or 0B in even dimensions, 
depending on the choice of chirality in the GSO projection. 
In odd space-time dimensions there is only one such theory due to the lack of
chirality. We shall refer to it as type 0AB.

Although the full conformal field theory corresponding to the above 
non-critical string theory has not yet been constructed, there are  
some indications that such a construction is indeed possible:

Consider the issue
of modular invariance. Define the fermionic traces for the pair of
Majorana--Weyl fermions $\psi(z)$, $\bar\psi(z)$ as\footnote{We use the 
notation of~\cite{Polchinski:a}. Here $q=\exp(2\pi i \tau)$.}
\beqs
Z^0_0(\tau) &=& \tr_{\mathrm NS} \bigg(q^N\bigg) \nonumber \\
Z^0_1(\tau) &=& \tr_{\mathrm NS} \bigg((-)^F q^N\bigg) \nonumber \\
Z^1_0(\tau) &=& \tr_{\mathrm R} \bigg(q^N\bigg) \nonumber \\
Z^1_1(\tau) &=& \tr_{\mathrm R} \bigg((-)^F q^N\bigg). \label{zetas}
\eeqs
In the ordinary type II string theory in $d$ dimensions,
modular invariance requires that the holomorphic contribution of the 
fermions to the partition function
\beq
Z^0_0(\tau)^{(d-2)/2} - Z^0_1(\tau)^{(d-2)/2} - Z^1_0(\tau)^{(d-2)/2} 
\pm Z^1_1(\tau)^{(d-2)/2}
\eeq
and the analogous antiholomorphic contribution be separately modular 
invariant up to an overall opposite phase.
It is well known that the lowest dimension where the product of the 
holomorphic contribution and the antiholomorphic contribution is modular
invariant is $d=10$.

In type 0 theories, the joined contributions of the holomorphic and the
antiholomorphic sectors give rise to 
\beq
|Z^0_0(\tau)|^{d-2} + |Z^0_1(\tau)|^{d-2} + |Z^1_0(\tau)|^{d-2} 
\pm |Z^1_1(\tau)|^{d-2},
\eeq
which is modular invariant for any $d$. Of course, for $d\not=10$ the explicit
expressions for the fermionic traces will be modified because of the 
changes in the spectrum, but we view the above as an indication that the
continuation ``off-criticality'' is more likely to work for the type 0 
string than for the usual type II.

Another objection that needs to be addressed is the ``$d=2$ barrier''. 
Let us briefly recall the physics behind this problem as presented 
e.g.~in~\cite{Polchinski:a}. Because
we do not yet know how to describe RR fields at the level of the sigma-model
we are forced to discuss this argument in the case of the bosonic theory. In
this context it is well know that there exists an exact CFT solution in
any $d$ in which the only non-zero background fields are a flat metric
$g_{MN}$, a 
linearly rising dilaton $\Phi$ and an exponential tachyon $T$
\beq
g_{MN}=\eta_{MN}, \quad\Phi=\sqrt{\frac{26-d}{6}}X^1, \quad
T=\exp\left(\left( \sqrt{\frac{26-d}{6}}-\
\sqrt{\frac{2-d}{6}}\right)X^1 \right).
\eeq

For $d\leq 2$ the background tachyon is
exponentially rising, preventing the string from entering the region of 
strong coupling. Moreover, fluctuations around the tachyon background are 
stable and thus the theory is well defined. On the contrary, for $d>2$, the
background tachyon oscillates. In principle this could still act as a cutoff
for the string coupling but the fluctuations have some negative frequency 
square modes and the theory becomes unstable.

As stressed in~\cite{Polyakov:a, Polyakov:b} we should not think of the 
``$d=2$ barrier'' as a no-go theorem but rather as an indication that
solutions for $d>2$ will necessarily involve a curved space-time metric.
This is the type of situation that is of interest in the connection
with gauge theory so, in a sense, it is to be expected that flat space-time
be ruled out. Unfortunately, we 
do not yet have an example of an exact CFT of this type and we
are forced to work order by order in $\alpha^\prime$ at the level of the
effective action. 
But it should be clear that there are no a priori reasons
for why there should not exist an exact solution.

A third encouraging sign comes from the analysis of the RR sector performed
in Section 3. By making some plausible assumptions about the massless
degrees of freedom it is possible to construct a rather compelling picture
of the RR sectors in various dimensions and their couplings, including
Chern--Simons terms. For instance, the necessity of doubling the RR spectrum 
in $d=4$ or $d=8$ Minkowski space-time is seen as coming from the fact 
that there are no  real self-dual forms in these dimensions.

Finally, let us note that considering $d<10$ from the sigma model point of
view is a very natural thing to do if the string theory has a 
perturbative tachyon.
The only way for a theory with a perturbative tachyon to make sense is if
there exists a mechanism through which the tachyon field condenses by acquiring
a vacuum expectation value. The tachyon potential at that point will then
give rise to a tree-level contribution to the cosmological constant by 
shifting the central charge. Since the effective central charge is going to 
be different from zero anyway\footnote{It seems unnatural and there is no 
symmetry argument for which the tachyon potential should vanish at that 
point.}, one 
is led to consider the theory with the most general value for the 
effective central charge
\beq
    c_{\rm eff.} = 10 - d - \frac{1}{2}V\bigg(\langle T \rangle\bigg).
\eeq 
{}From the target space point of view this acts as a contribution to the 
cosmological constant and thus shows that, for $d>2$, one should look at 
curved space-times.

None of the above points constitute a proof that conformally invariant 
solutions to the type 0A/B string exist for arbitrary $d$ but we view them
as strong indications that such construction is possible. Having taken this 
as our basic assumption, throughout the paper we work at the level of the
effective action to one loop in $\alpha^\prime$, i.e. the gravity level
without higher order corrections. 

The form of the action of the type 0 gravity, can be determined perturbatively,
with a certain number of ambiguities involving the tachyon potential and
the coupling to the RR fields, which are essentially there because of lack
of supersymmetry on the target space. However, supersymmetry on the world
sheet allows one to make some statements on these terms, 
as we will see in Section 3. We will stay generic and will 
be able to show that in any
dimension there exists a set
of exact solutions of the classical equations of motion, which give AdS metric
and involve a non-zero RR field, other than constant dilaton and tachyon. 
Such solutions depend only on a {\it finite} number of parameters for which
a string-theoretical derivation is still lacking. 

The solutions found represent Polyakov's conformal fixed points in the dual
gauge theory --- they support a condensed tachyon, but the first issue to be
addressed is the stability against quantum fluctuations of the fields. 
Because of the mixing of several of these fields it is not enough to 
analyze the tachyon itself, one has to disentangle the full set of
fluctuations~\cite{Ferretti:b}. For simplicity, we restrict
to the case where the space is $d$-dimensional AdS, so that there are no
KK modes to be worried about. There the analysis simplifies 
considerably, but it is
still non trivial because of dilaton-tachyon mixing.

Another issue that was raised in~\cite{Polyakov:a} is the field 
theory interpretation of these solutions. There it was claimed that they
may represent an interacting UV conformal point. 
To address this issue one needs to show that these gravity 
solutions represent in fact a point in wider space, that is actually the
RG phase diagram of the field theories we have at hand. 
In this enlarged theory space there may be more than one conformal solution 
and there exist trajectories which interpolate between these 
points. 
 
These considerations have already been realized in the framework of 
type IIB supergravity, where the RG flow is believed 
to be driven by operators dual to scalar fields coming 
from KK reduction in the compact 
directions~\cite{Distler, Girardello:a, Girardello:b, Porrati}, or
by the running of the dilaton \cite{Kehagias:a, Gubser:b, deMelloKoch,
Kehagias:b}. It turns out that
type 0 gravity provides an example of this general feature
of the AdS/CFT correspondence, as well. 
In fact, the physics is already captured
by the set of solutions involving no compact space. Even without including 
KK modes, the tachyon will mix with the dilaton field, and generate 
on the field theory side a RG flow that connects interacting 
conformal fixed points. 

\section{A look at the type 0A/B theory in $d=10$}

In this section we briefly summarize some known facts about the perturbative
properties of these theories. We will thus set $d=10$ throughout this section
and consider perturbation theory around the (unstable) vacuum. We will, 
however, point out the 
various places where modifications occur when one is 
considering $d\not= 10$. We will return to these changes in  Section 3
where we present the detailed structure of the RR terms in $d\leq10$.
 
In the notation of~\cite{Polchinski:a}, there are (up to equivalences) 
only four consistent ways of
combining the various sectors of the ${\cal N}=(1,1)$ NSR closed oriented 
string in $d=10$:
\beqs
    {\mathbf IIA}\;\;&:&\;\;
    (NS+,NS+)\oplus(NS+,R+)\oplus(R-,NS+)\oplus(R+,R-)\nonumber \\
    {\mathbf IIB}\;\;&:&\;\;
    (NS+,NS+)\oplus(NS+,R+)\oplus(R+,NS+)\oplus(R+,R+)\nonumber \\
    {\mathbf 0A}\;\;&:&\;\;
    (NS+,NS+)\oplus(NS-,NS-)\oplus(R+,R-)\oplus(R-,R+)\nonumber \\
    {\mathbf 0B}\;\;&:&\;\;
    (NS+,NS+)\oplus(NS-,NS-)\oplus(R+,R+)\oplus(R-,R-). \label{GSO}
\eeqs
The first two are the usual type IIA/B superstring and the last two are 
those of interest here. The massless fields of the theory are the dilaton
$\Phi$, graviton $g_{MN}$ and two-form $B_{MN}$ coming from the $(NS+,NS+)$
sector and twice as many RR fields coming from the doubled RR sectors.
There is a tachyon $T$ from the $(NS-,NS-)$ and no fermionic mode at all. All 
other modes are massive. 

\subsection{Selection rules for $d=10$}

The lack of space-time fermions means that the 
theory has no space-time supersymmetry; however, the presence of 
${\cal N}=(1,1)$ world-sheet supersymmetry has the
following simplifying features~\cite{Klebanov:a}:

\begin{itemize}
\item[a)] All tree-level correlators involving an odd number of tachyons
and only $(NS+, NS+)$ vertex operators are zero. This can be seen from the 
explicit form of the vertex operators. 
The vertex operator for the massless $(NS+, NS+)$ states and
for the tachyon in the $(NS-, NS-)$ sector are, again in the
notation of~\cite{Polchinski:a}
\beqs
V^{0,0}_{(NS+, NS+)}&=&-(i\partial X^M +k\cdot\psi~ \psi^M)
       (i\bar\partial X^N +k\cdot\tilde\psi~ \tilde\psi^N)~
       \ee^{ik\cdot X}\nonumber\\
V^{-1,-1}_{(NS+, NS+)}&=&\ee^{-\phi-\tilde\phi}~
                  \psi^M\tilde\psi^N~ 
\ee^{ik\cdot X}\nonumber\\
V^{0,0}_{(NS-, NS-)}&=&k\cdot\psi~ k\cdot\tilde\psi~ 
\ee^{ik\cdot X}\nonumber\\
V^{-1,-1}_{(NS-, NS-)}&=&\ee^{-\phi-\tilde\phi}~ 
\ee^{ik\cdot X}.\label{nsvertex}
\eeqs
Since on the sphere we need to take any two of the above in the $(-1,-1)$ 
picture and all the rest in the $(0,0)$ picture it is clear that we will always
end up with the correlation function of an odd number of $\psi$'s 
which vanishes.
In particular, notice that in this vacuum $\langle T \rangle = 0$ whereas
we expect a tachyon condensate in the true vacuum. 
\item[b)] All tree-level correlators involving an odd number of tachyons,
any number of fields from the $(NS+, NS+)$ and two RR fields from the
same sector must vanish.
For instance, if the amplitude with one tachyon did not vanish there
would be a tachyon pole in some tree-level correlation 
function of the corresponding type II theory. The vanishing of the 
amplitude can also be seen at the level of the correlation functions of the
vertex operators. We can always use the vertex operators for the RR fields
in the $(-1/2, -1/2)$ picture -- written as a bispinor it reads
\beq
    V^{-1/2,-1/2}_{(R\pm, R\pm)}=\ee^{-\phi/2-\tilde\phi/2}~ 
    \Theta_\alpha\tilde\Theta_\beta~ \ee^{ik\cdot X}~. \label{rrvertex}
\eeq
In a correlator that involves  
two such vertex operators, one remaining $(NS\pm,NS\pm)$
vertex (\ref{nsvertex}) needs to be taken in the $(-1,-1)$ picture yielding
an overall even number of $\psi$'s. Schematically, this part of the correlator
is proportional to $C\Gamma^{2n}$ and only connects spinors of opposite 
chirality because the charge conjugation matrix $C$ anticommutes with the
chirality matrix $\Gamma_\chi$. 
This is also true in $d=6$ whereas in 
$d=4$ and $d=8$ the opposite is true, since now the matrix $C$ commutes with 
$\Gamma_\chi$. Hence, in $d=4$ and $d=8$ we expect to find a coupling between 
an odd number of tachyons and RR fields from the same sector and none 
between an odd number of tachyons and RR fields from the opposite sector. 
\item[c)] The reverse statement holds for an even number of 
tachyons (in particular 
no tachyon at all). The correlators will now involve an odd number of $\psi$'s
and these vanish in $d=10$ between spinors of 
opposite chirality. Again, it is
natural to expect that such a statement will hold in $d=6$ and be reversed in
$d=4$ and $d=8$.
\end{itemize}

For odd space-time dimensions, both statements b) and c) 
are empty due to the lack of chirality and we do not expect any particular 
symmetry of the couplings.
Statement c) is also consistent with the fact that in $d=10$ 
all correlation functions
between $(NS+, NS+)$ fields and fields from \emph{one} given RR sector are
the same as in type II theory, which is obvious because the vertex operators 
and their correlation functions are precisely the same and there is no 
loop in which other modes could propagate.
In particular, it indicates that there are Chern--Simons terms in the 
effective action and that the various field strengths need to be modified 
by shifting them with a $B_{MN}$ dependent transformation just as in
type II supergravity. It seems natural to postulate that such terms will
be present also in lower dimensions in the cases 
where they are allowed by the symmetries of the problem.

These simple properties and some explicit tree-level 
computations~\cite{Klebanov:a} allow one 
to write down, for the critical case, to first non-trivial
order in $\alpha^\prime$, an effective action up to terms quadratic in the
gauge fields. 
Below we shall present our proposed generalization of this action in 
any dimension including Chern--Simons terms.

\subsection{D-branes in the $d=10$ theory}

We conclude this section with a look at the type of D-branes in this 
theory~\cite{Bergman, Billo}. In this subsection we always work in $d=10$.

Due to the doubling of the RR fields, there are twice as many D-branes as 
in the corresponding type II theory. Let us denote 
by $F$ and $F^\prime$ the RR fields from the
two RR sectors of (\ref{GSO}) respectively\footnote{We refrain from using the
notation $F$ and $\bar F$ used in~\cite{Klebanov:a} because we reserve such
notation for the $d=4$ and $d=8$ case where such fields are complex 
conjugate of each other.}. For a given $p$ we have thus
four types of elementary branes (counting the anti-branes) with charges
$(Q=1, Q^\prime=1)$, $(Q=1, Q^\prime=-1)$, $(Q=-1, Q^\prime=1)$,
$(Q=-1, Q^\prime=-1)$. Branes charged only with respect to, say, $F$ are 
possible but carry charge $(Q=2n, Q^\prime=0)$ in these units and thus can be 
built from the four constituents above.

A quick way to understand the spectrum of massless excitations living on 
the world-volume of a stack of such branes is to consider the closed string 
exchange between two such branes, perform a modular transformation and
read off the spectrum from the open string sector.

Let us denote the usual open string traces in the various sectors 
as\footnote{We use the symbol $\Tr$ to denote the sum over all eight 
transverse bosonic and fermionic components, to distinguish it from the
trace $\tr$ in (\ref{zetas}).}
\beqs
\Tr_{\mathrm NS} \bigg(q^N\bigg) &=& 
        \left(\frac{f_3(q)}{f_1(q)} \right)^8\nonumber \\
\Tr_{\mathrm NS} \bigg((-)^F q^N\bigg) &=&
        \left(\frac{f_4(q)}{f_1(q)} \right)^8 \nonumber \\
\Tr_{\mathrm R} \bigg(q^N\bigg) &=& 
        \left(\frac{f_2(q)}{f_1(q)} \right)^8,
\eeqs
where, as usual,
\beqs
    f_1(q) &=& q^{1/12}\Pi(1-q^{2n}) \nonumber \\
    f_2(q) &=& \sqrt{2} q^{1/12}\Pi(1+q^{2n}) \nonumber \\
    f_3(q) &=& q^{-1/24}\Pi(1+q^{2n-1}) \nonumber \\
    f_4(q) &=& q^{-1/24}\Pi(1-q^{2n-1}). 
\eeqs

The closed string exchange can be written in terms of these functions by 
making a modular transformation $\tilde q=\ee^{-\pi\tilde t}\to 
q=\ee^{-\pi /\tilde t} = \ee^{-\pi t}$.
Let us denote by $H$ the light-cone oscillator part of the closed string
Hamiltonian, and use the boundary states $|B\rangle$ to denote the branes
(cf.~\cite{Hussain, DiVecchia} for a complete discussion of the boundary
state formalism). Let us also introduce the shorthand notation
\beq
    \left[NS+, NS+\right]
    =\langle B|\ee^{-\tilde t H}|B\rangle_{(NS+, NS+)},
\eeq
and similar expressions for the other sectors. Then, after a modular 
transformation:
\beqs
    \left[NS+, NS+\right]&\to& \frac{1}{2}
    \left(\Tr_{\mathrm NS} \bigg(q^N\bigg) -
          \Tr_{\mathrm R} \bigg(q^N\bigg) \right) = \frac{1}{2}
    \left(\frac{f_3(q)^8 - f_2(q)^8}{f_1(q)^8} \right) \nonumber\\
    \left[NS-, NS-\right] &\to& \frac{1}{2}
    \left(\Tr_{\mathrm NS} \bigg(q^N\bigg) +
          \Tr_{\mathrm R} \bigg(q^N\bigg) \right) = \frac{1}{2}
    \left(\frac{f_3(q)^8 + f_2(q)^8}{f_1(q)^8} \right) \nonumber\\
    \left[R\pm, R\pm\right] &\to& -\frac{1}{2}
          \Tr_{\mathrm NS} \bigg((-)^F q^N\bigg) = -\frac{1}{2}
    \left(\frac{f_4(q)^8}{f_1(q)^8} \right).\\
\eeqs
All RR sectors give the same contribution.

There are three elementary cases to consider. First, consider the case of
two like-like charged branes $(Q_1=1, Q^\prime_1=1)$ and 
$(Q_2=1, Q^\prime_2=1)$.
Let us consider the type 0B case for definitiveness; all considerations apply 
to the type 0A case as well.
In this case we have the following situation:
\beq
    \left[NS+, NS+\right]+\left[NS-, NS-\right]+
    \left[R+, R+\right]+
    \left[R-, R-\right] \to
    \frac{f_3(q)^8 - f_4(q)^8}{f_1(q)^8}. 
    \label{electric}
\eeq
{}From (\ref{electric}) we read off, with the wisdom 
of~\cite{Polchinski:b, Witten:b},
that the world-sheet theory has no tachyon,
no fermions, and the same massless bosons as pure $d=10$ Yang-Mills theory
dimensionally reduced to $p+1$ dimensions. 

Second, consider the case  $(Q_1=1, Q^\prime_1=1)$ and 
$(Q_2=1, Q^\prime_2=-1)$.
Here the exchange is
\beq
    \left[NS+, NS+\right]-\left[NS-, NS-\right]+
    \left[R+, R+\right]-
    \left[R-, R-\right] \to
    -\frac{f_2(q)^8}{f_1(q)^8}. \label{fermi}
\eeq
Perhaps the only subtlety is the minus sign in front of the $(NS-, NS-)$ 
exchange. This sign is fixed by looking at the coupling of the tachyon
to the branes and noticing~\cite{Klebanov:a} that it has a coupling constant
proportional to the product of the charges of the brane, thus yielding
$Q_1 Q^\prime_1 \times Q_2 Q^\prime_2 = -1$.
{}From (\ref{fermi}) we see that this brane configuration has only fermions
on the world-volume. These two configurations were  
studied in~\cite{Klebanov:c}.

Finally the brane/anti-brane case is given by
$(Q_1=1, Q^\prime_1=1)$ and $(Q_2=-1, Q^\prime_2=-1)$ and corresponds to
\beq
    \left[NS+, NS+\right]+\left[NS-, NS-\right]-
    \left[R+, R+\right]-
    \left[R-, R-\right] \to 
    \frac{f_3(q)^8+f_4(q)^8}{f_1(q)^8}.
\eeq
There is an open string tachyon in this theory, just as in~\cite{Green, Banks},
signaling an instability of the theory.

\section{Type 0 effective actions, Ramond--Ramond fields 
and Chern--Simons terms}

In this section we present our proposal for the effective action of type 0
string theory in any dimension including the Chern--Simons couplings.
Lacking a formulation from ``prime principles'', the identification of the
RR sectors and their couplings requires a certain amount of guesswork. The 
picture that emerges, however, is quite simple and satisfying. We shall
see, for instance, that 
it gives support to the idea that the RR sectors must be doubled compared 
to the type II string.

\subsection{The NS-NS sector}

The NS-NS sector is common to all of these theories and can in principle be
obtained from a sigma-model approach. It involves the massless fields of the
$(NS+,NS+)$ sector (a dilaton $\Phi$, a graviton $g_{MN}$ and an 
antisymmetric tensor $B_{MN}$) 
and a tachyon $T$ from the $(NS-,NS-)$ sector. 
The tachyon potential $V(T)$ is an even function from the property a)
of the previous section~\cite{Klebanov:a}. The relevant action is thus,
in the string frame
\beq
S_{NS-NS} = \int \dd^dx\;\sqrt{-g}\left\{\ee^{-2\Phi}\left(
R -\frac{1}{12} |\dd B|^2 + 4 |\dd\Phi|^2 - 
\frac{1}{2}|\dd T|^2 - V(T) \right) \right\}, \label{nspart}
\eeq
where it is natural to absorb the central charge deficit $10-d$ into the 
definition of the tachyon potential, i.e.
\beq
      V(T) = -10 + d - \frac{d-2}{8} T^2 + \cdots
\eeq

It should be kept in mind that (\ref{nspart}) is by no means unique. It 
suffers from the usual ambiguities that come from extrapolating on-shell
data. In particular, there could be arbitrary (even) functions of $T$
multiplying the various kinetic terms in the Lagrangian. Up to this order in 
$\alpha^\prime$ this is essentially all that can happen\footnote{Things 
become even 
more complex at the next order in $\alpha^\prime$ -- for instance there
could be terms of the type $T^{2n}R^{MN}\partial_M T\partial_N T$.}.
As shown in the Appendix
A of~\cite{Klebanov:a} however, precisely because of their ambiguous nature 
it is possible to redefine away some of them, such as the term $RT^2$. At the
same time, terms of the type $T^2 |\dd B|^2$ and their counterpart
for the RR kinetic terms are needed and should be kept.
In the following we will never need the field $B_{MN}$ and we will 
set the coefficients of $R$, $|\dd\Phi|^2$ and $|\dd T|^2$ 
as in (\ref{nspart}),
the main conclusions being independent of the presence of such terms.

\subsection{RR kinetic and Chern--Simons terms}

One guiding principle~\cite{Armoni} in the identification of the RR sector 
is the idea that for any $d$ there will still be massless 
excitations in the R sector of the open string and thus their on-shell
degrees of freedom will fall into representations of the little group
$SO(d-2)$. The resulting situation is best summarized in the table below. 

In ten dimensions one obtains massless RR fields in the type 0A theory by 
considering tensor products of spinors of different chiralities 
$(\mathbf{+-})$ and $(\mathbf{-+})$, and in the 0B theory of same 
chiralities $(\mathbf{++})$ and $(\mathbf{--})$. This can be readily 
generalized for any non-critical even dimension, whereas it is not 
quite clear what the right generalization to odd dimensions is. Note, 
however, that an odd dimensional bispinor can be decomposed in terms 
of the lower even dimensional bispinors as 
$({\mathbf ++})\oplus({\mathbf +-})\oplus({\mathbf -+})\oplus({\mathbf --})$, 
and that this  is the sum of the field contents of the 
0A and the 0B theories of one lower dimension.
It thus seems reasonable to assume that a sum of two modular invariant 
sectors should yield a modular invariant theory in one dimension higher 
without doubling the RR spectrum by hand\footnote{A different point of 
view was taken in \cite{Armoni}. It will be interesting to find a resolution
to this puzzle.}.
 
{\footnotesize

\begin{center}
\begin{tabular}{c|c|c|c|c|c}
$d$ & $SO(d-2)$ & Spin reps. & & $R\times R$ sector(s)& 
                               Real off-shell fields \\
    &&&&&\\ \hline &&&&&\\
$4$ & $U(1)$ & ${\mathbf 1}_{1/2}$, ${\mathbf 1}_{-1/2}$ 
&0A& ${\mathbf 1}_0+{\mathbf1}_0$ & $2A$ \\
    &&&0B& ${\mathbf 1}_{-1}+{\mathbf 1}_{1}$ & $A_M$ \\
&&&&&\\ \hline &&&&&\\ 
$5$ & $SU(2)$ & $\mathbf{2}$ &0AB & $\mathbf{1}+\mathbf{3}$ & 
      $A, A_M$  \\
&&&&&\\ \hline &&&&&\\
$6$ & $SU(2)^2$ & $(\mathbf{1},\mathbf{2}),\; (\mathbf{2},\mathbf{1})$ &0A& 
$2(\mathbf{2},\mathbf{2})$ & 
$2A_M$ \\
    &&&0B& $2(\mathbf{1},\mathbf{1})+(\mathbf{1},\mathbf{3})+
(\mathbf{3},\mathbf{1})$ & 
$2A, A_{MN}$ \\
&&&&&\\ \hline &&&&&\\
$7$ & $ Sp(4)$ & $\mathbf{4}$ &0AB& $\mathbf{1}+\mathbf{5} +\mathbf{10}$ & 
      $A, A_M, A_{MN}$  \\
&&&&&\\ \hline &&&&&\\
$8$ & $SU(4)$ & $\mathbf{4},\; \mathbf{\bar 4}$ &0A&
     $2(\mathbf{1}+\mathbf{15})$ & $2A, 2A_{MN}$ \\
 &  & &0B& $\mathbf{6}+\mathbf{10}+\mathbf{\bar{6}}+\mathbf{\bar{10}} 
$ & $2A_M, A_{MNR}$\\
&&&&&\\ \hline &&&&&\\
$9$ & $SO(7)$ & $\mathbf{8}$ &0AB& 
$ \mathbf{1}+\mathbf{7}+\mathbf{21}+\mathbf{35}$ & 
      $A, A_M, A_{MN}, A_{MNR}$ \\
&&&&&\\ \hline &&&&&\\
$10$& $SO(8)$ & $\mathbf{8}_s,\; \mathbf{8}_c$ &0A& 
      $2(\mathbf{8}+\mathbf{56})$ & 
      $2A_M, 2A_{MNR}$ \\
 &  & &0B& $2(\mathbf{1}+\mathbf{28}+\mathbf{35})$ & 
$2A, 2A_{MN}, A_{MNRP}$
\end{tabular}
\end{center}

}

To understand the table consider for example $d=8$. In the type 0B theory,
the bispinor in the $(R+,R+)$ sector decomposes into 
$\mathbf{4}\times\mathbf{4}=\mathbf{6}+\mathbf{10}$. These 
are all complex representations that yield a complex vector and a complex
three form with a self dual field strength. Notice that it is possible to 
construct a self dual form in $d=8$ only if it is complex because $*^2=-1$.
The bispinor in the  $(R-,R-)$ sector yields the complex conjugate fields.
These two sets of fields can be combined into two real one-forms and one real
three-form without any duality constraint. These are the fields written 
in the last column.

Notice that in odd dimensions we have only one version of the theory 
(0AB) without
the restriction on the rank of the forms. 
In even dimensions the forms 
come in even or odd rank depending on the RR projection. In $d=10$ and 
$d=6$ the assignment is the familiar one (odd forms for type A and even 
for type B) whereas in $d=8$ and $d=4$ it is reversed. Of course, it is 
possible to dualize some of the fields to obtain the magnetically
charged branes. The unique form of degree $d/2 -1$ for the type 0B case  
admits both electric and magnetic charges.

Notice that if it were not for the doubling of the RR sectors it would be 
impossible to write off-shell real fields for the 0B theory 
in $d=4$ and $d=8$
because, due to the Minkowski signature, it is impossible to impose either 
self-duality or anti-self-duality on real forms. 
We view this fact, together with the
argument based on modular invariance, as yet another piece of evidence 
for the necessity of the presence of both RR sectors.

To obtain the complete form of the RR couplings up to two derivatives we 
need to address the issue of Chern--Simons terms. The terms of relevance here 
are those constructed with 3 gauge fields and 2 derivatives.
Despite the lack of space-time supersymmetry, 
the Chern--Simons terms are present, 
at least in $d=10$, because they are there in the type II theories.
It thus seems
that the presence of these terms is dictated more by world-sheet supersymmetry
than by space-time supersymmetry and it is natural 
to assume that such terms are also present in lower dimensions.
The presence of so many RR fields may seem to lead to difficulties in
determining these terms. 
However, there are two simplifying features that we infer from 
the $d=10$ case: First, there will not be terms involving only RR fields
since they correspond to the correlation function of an odd number of 
spin fields. One of the three gauge fields must therefore be the 
$(NS+, NS+)$ two-form $B_{MN}$. Second, applying the selection rules of the
previous section, we see that the coupling will involve fields from the same
RR sector in $d=6$ and $d=10$ and from the opposite sector in $d=4$ and $d=8$.

Let us start with the case of $d$ odd. In this case our proposal for the
RR part of the action is
\beqs
    S^{{\mathrm 0AB}}_{d=5} &=& \int\;\ff 
(F_1*F_1 + F_2*F_2) + B F_1 F_2\nonumber\\
    S^{{\mathrm 0AB}}_{d=7} &=& \int\;\ff 
(F_1*F_1 + F_2*F_2 + \tilde F_3*\tilde F_3) 
   + B F_2 F_3\nonumber\\
    S^{{\mathrm 0AB}}_{d=9} &=& \int\;\ff (F_1*F_1 + F_2*F_2 + 
    \tilde F_3*\tilde F_3 + \tilde F_4*\tilde F_4) 
   + B F_3 F_4. \label{actodd}
\eeqs
The forms $F_n$ are the field strengths associated to the RR gauge potentials.
In this section we use ``index free'' notation and redefine the normalization 
coefficients to one in order not to clutter the formulas too much. A wedge
product between forms is always understood. The tilde above the forms always
indicates the ``NS-NS shift'', e.g. $\tilde F_4=F_4 + B F_2$, with the 
appropriate modified gauge transformation just as in type II supergravity. 
Notice that, once it is assumed that the Chern--Simons terms are present, the
modification in the field strength must also be present for the action to 
transform correctly under electric/magnetic duality. $f(T)$ is a function of 
the tachyon whose first few coefficients in the Taylor expansion around zero 
can in principle be determined by extrapolating from the on-shell 
computation~\cite{Klebanov:a} in $d=10$. We shall see that the detailed form 
of these functions is not directly relevant for the computations of the 
properties of the dual field theories.

To write the actions for the even dimensional cases, let us denote the field
strengths from the two RR sectors by $F, \; F^\prime$ for $d=6$ and $d=10$
and by $F, \; \bar F$ for $d=4$ and $d=8$. In the former case the field
strengths are real whereas in the latter they are complex conjugates of each 
other. The form of highest degree in the type 0B case is special -- it is 
self dual in the complex case and its vertex operator does not contain the
chiral projection. 

{
\beqs
S^{{\mathrm 0A}}_{d=4} &=& \int\;\fe (F_1*\bar F_1) + 
\fo (F_1*F_1+ \bar F_1*\bar F_1)
               +i B F_1 \bar F_1\nonumber \\
S^{{\mathrm 0B}}_{d=4} &=& \int\;\fe (F_2*\bar F_2) + 
\fo (F_2*F_2 +\bar F_2*\bar F_2)
               \nonumber \\
S^{{\mathrm 0A}}_{d=6} &=& \int\;\fe (F_2* F_2 +  F^\prime_2* F^\prime_2) 
               + \fo (F_2* F^\prime_2) + B (F_2 F_2 +  F^\prime_2 F^\prime_2)
               \nonumber \\
S^{{\mathrm 0B}}_{d=6} &=& \int\;\fe (F_1* F_1 +  F^\prime_1* F^\prime_1+
               \tilde F_3 *\tilde F_3 ) 
               + \fo (F_1* F^\prime_1 + \tilde F_3 *\tilde F_3 ) \nonumber\\
               && + B (F_1 F_3 +  F^\prime_1 F_3)
               \nonumber \\
S^{{\mathrm 0A}}_{d=8} &=& \int\;\fe (F_1*\bar F_1 + 
\tilde F_3 * \tilde {\bar F_3})
               +\fo (F_1*F_1+ \bar F_1*\bar F_1 + \tilde F_3 * \tilde F_3
               \nonumber \\ &&+\tilde{\bar F_3} * \tilde{\bar F_3} )
               +i B F_3 \bar F_3\nonumber \\
S^{{\mathrm 0B}}_{d=8} &=& \int\;\fe (F_2*\bar F_2 + \tilde F_4 * 
\tilde{\bar F_4})
               +\fo (F_2*F_2 +\bar F_2*\bar F_2 \nonumber\\
               &&+ \tilde F_4 * \tilde F_4
               +\tilde{\bar F_4} * \tilde{\bar F_4})
               + B (F_2 \bar F_4 + F_4  \bar F_2)
                \nonumber \\
S^{{\mathrm 0A}}_{d=10} &=& \int\;\fe (F_2* F_2 +  F^\prime_2* F^\prime_2 +
               \tilde F_4* \tilde F_4 +\tilde F^\prime_4*\tilde F^\prime_4) 
               \nonumber \\&&+ \fo( F_2* F^\prime_2 + 
               \tilde F_4*\tilde F^\prime_4)
               + B (F_4 F_4 +  F^\prime_4 F^\prime_4)
               \nonumber \\
S^{{\mathrm 0B}}_{d=10} &=& \int\;\fe (F_1* F_1 +  F^\prime_1* F^\prime_1+
               \tilde F_3* \tilde F_3 +  \tilde F^\prime_3* \tilde F^\prime_3
               + \tilde F_5 *\tilde F_5 )\nonumber \\&& 
               + \fo (F_1* F^\prime_1 + \tilde F_3 *\tilde F^\prime_3
               + \tilde F_5 *\tilde F_5 ) 
                + B (F_3 F_5 +  F^\prime_3 F_5).\label{acteven}
\eeqs
}
We present the form of the actions in all their generality because it will be 
useful for future more detailed computations. For our present purposes 
however, it should be noticed that the kinetic terms in the actions can be
diagonalized by letting $F_\pm = F\pm F^\prime$ in $d=6,10$ and
$F_\pm = F\pm i\bar F$ in $d=4,8$.

\subsection{Massive type 0 gravity}

There is still one RR form field that can be added to the actions 
(\ref{actodd}) and (\ref{acteven}). 
In a $d$-dimensional space-time it is possible to 
introduce a rank $d-1$ gauge potential coupling to a corresponding
extended object. It carries no physical degrees of freedom and
therefore it is not visible in the on-shell analysis of the previous 
subsection. Its rank $d$ field strength, however, carries an energy density
and it does affect the physics. This form will be used in Section 5
when constructing various field theory duals. This case will provide 
the simplest example which displays  
most of the interesting physics, and it allows one to avoid 
the complications of disentangling the Kaluza-Klein 
modes.

In type IIA supergravity 
(and thus in $d=10$ type 0A for each RR sector) it is well known
how to introduce such a field~\cite{Romans,Bergshoeff:1996ui}. 
The required modifications in the bosonic sector are the addition of the 
terms
\beqs
     \int\; M F_{10} +\frac{1}{2} M^2 *1 \label{addit}
\eeqs
to the action, and a further shift of the 2- and the 4-form 
field strengths by
$MB$ and $MB^2/2$, respectively. The gauge transformations are changed 
accordingly in order to re-ensure gauge invariance. Integrating over 
the gauge potential of $F_{10}$ imposes the constraint that $M$ be constant. 
Solving the equation of motion for $M$ establishes a connection 
between $M$ and $F_{10}$. In the case $B=0$ -- relevant to our analysis --  
they are simply Hodge duals of each other as is readily seen from 
(\ref{addit}).

From the string theory point of view \cite{Polchinski:b}, 
the natural generalization of the RR $\beta$-function 
equations implies $\dd * F_{d}=0 $ and $\dd F_{d}=0$, 
as the top-form, too, appears in the reduction of 
an even dimensional type 0A 
bispinor into antisymmetric tensor representations.
In our case, we must also include the coupling with the tachyon. 
The relevant addition is 
\beqs
- \frac{1}{2} \int f(T) F_{d} * F_{d} 
\eeqs
that we assume be present in any dimension.

\section{Classical solutions}

In what follows we shall show that the above described low energy 
theories allow Freund--Rubin type solutions~\cite{Freund}, where the 
dilaton and the 
tachyon are constant, 
the space-time factorizes into a product of an AdS space and a sphere, 
and the only nontrivial 
form-field is a RR field. Such types of solutions are very familiar from the
supergravity literature, see e.g.~\cite{Duff:a, Duff:b, Kim, 
Salam, Randjbar-Daemi}.

It is sufficient to consider the Einstein frame action\footnote{We now switch
to component notation for clarity and reinstate all the appropriate 
normalizations. Note that our normalization of the tachyon differs by a factor
of $\sqrt{2}$ from that of most of the recent literature.}
\beqs
S &=& \int \dd^dx \sqrt{-g}~ \Bigg\{ R - \frac{1}{2} (\partial_{M}\Phi)^2 
- \frac{1}{2} (\partial_{M}T)^2 - V(T)~\ee^{a\Phi} \nonumber \\ & & \qquad
- \frac{1}{2~(p+2)!}~ f(T)~ \ee^{b\Phi}~ \Big(F_{M_1 \cdots M_{p+2}}\Big)^2  
\Bigg\},
\eeqs
where $V(T)$ is the sum of the tachyon potential and the central 
charge deficit, 
and $f(T)$ is  the  coupling between the $(p+2)$-dimensional RR 
form $F$ and the tachyon. The RR gauge field is the appropriate linear 
combination of some of the fields of the previous section in such a way that
the kinetic terms are diagonal. After diagonalization, $f(T)$ no longer has 
any particular symmetry property.

The coefficients $a$ and $b$ are
\beqs
a &=& \sqrt{\frac{2}{d-2}} \\
b &=& \frac{1}{2} (d-2p-4) \sqrt{\frac{2}{d-2}} .
\eeqs
The field $B_{MN}$ that we are setting to zero 
here may appear linearly in the full 
action only 
in the Chern--Simons term, but in that case multiplied by  $F \wedge F$, 
which will vanish in the Freund--Rubin ansatz.

The equations of motion can be summarized as follows:
\beqs
\Box \Phi &=&  aV(T)~ \ee^{a\Phi} + \frac{b}{2}~ f(T)~ 
\ee^{b\Phi}~ \frac{1}{(p+2)!}~ \Big(F_{M_1 \cdots M_{p+2}}\Big)^2 
\label{dilaton} \\
\Box T &=& V'(T)~ \ee^{a\Phi} + \frac{1}{2}~ f'(T)~  
\ee^{b\Phi}~ \frac{1}{(p+2)!}~ \Big(F_{M_1 \cdots M_{p+2}}\Big)^2 
\label{tachyon}  \\
R_{MN} &=&  \frac{1}{2}~ \partial_M \Phi~ \partial_N \Phi +
\frac{1}{2}~ \partial_M T~ \partial_N T  \nonumber \\ 
& & + \frac{1}{d-2}~ g_{MN}~ V(T)~ \ee^{a\Phi} + 
\frac{1}{2}~ f(T)~ \ee^{b\Phi}~ \tilde{T}_{MN} \label{gravity} \\
0 &=&  \nabla^N~ \Big( f(T)~ \ee^{b\Phi}~ F_{NM_1 \cdots M_{p+1}}  
\Big)~. \label{formfield}
\eeqs
The tensor $\tilde T_{MN}$ is shorthand for the (trace subtracted) 
stress energy tensor
\beqs
\tilde{T}_{MN} &=& \frac{1}{(p+1)!} \Bigg( F_{MK_1 \cdots K_{p+1}}
 F_{N}^{~~K_1 
\cdots K_{p+1}} - \frac{(p+1)~ g_{MN}}{(p+2)(d-2)}  \Big(F_{K_1 
\cdots K_{p+2}}\Big)^2 \Bigg) \nonumber \\ 
& &
\eeqs
Again, we have ignored potential contributions from Chern--Simons terms 
because they vanish for the classical solution. They do contribute to the
analysis of the fluctuations in the general case and also for this reason,
in the next section, when computing the critical properties of the field 
theory duals we restrict to the simple case $d=p+2$ where such complications
do not arise. We hope to return to the general case in a later 
paper~\cite{Ferretti:b}.

These equations of motion have a solution with constant dilaton 
$\Phi=\Phi_0$ and tachyon $T=T_0$ in the gravity background of a product space 
\beqs
{\mathrm AdS}_{p+2} \times {\mathrm S}^{d-p-2}.
\eeqs
The size of the two maximally symmetric spaces is determined by 
setting\footnote{The Greek indices refer to the AdS space and the 
Latin indices to the sphere.}
(always in units of $\alpha^\prime$)
\beq
     R_{\mu\nu\rho\lambda}=-\frac{1}{R^2_0}\left(g_{\mu\rho}g_{\nu\lambda}
         -g_{\mu\lambda}g_{\nu\rho} \right)\quad
     R_{ijkl}=+\frac{1}{L^2_0}\left(g_{ik}g_{jl}-g_{il}g_{jk} 
         \right). \label{riemann}
\eeq
Finally, the RR field is set proportional to the volume-form 
of the anti-de Sitter 
space, and hence its only nontrivial components are 
\beqs
F_{\mu_1 \cdots \mu_{p+2}} &=& F_0~ \sqrt{-g({\mathrm AdS}_{p+2})}~ 
\epsilon_{\mu_1 \cdots \mu_{p+2}}, 
\eeqs
where the constant $F_0$ is related to the conserved charge $k$ by 
\beqs
k = f(T_0)~\ee^{b\Phi_0}~F_0~. \label{chargek}
\eeqs

Given the two functions $V(T)$ and $f(T)$, the tachyon and the dilaton vacuum 
expectation values are determined from Eqs.~(\ref{dilaton}), (\ref{tachyon}) 
and (\ref{chargek}). The tachyon $T_0$ 
can be expressed implicitly, as the solution of an algebraic equation, namely
\beqs
\frac{f'(T_0)}{f(T_0)} &=& \frac{1}{2}(d-2p-4)~ \frac{V'(T_0)}{V(T_0)}. 
\label{tnaught}
\eeqs
The dilaton $\Phi_0$ can then be readily obtained from
\beqs
\ee^{(a+b)\Phi_0} &=& \frac{(d-2p-4)}{4}~ \frac{k^2}{f(T_0)~V(T_0)}~.
\eeqs
The radii of the anti-de Sitter space $R_0$ and that of the sphere 
$L_0$ can be solved from the Einstein equations (\ref{gravity})
\beqs
R^2_0 &=& (p+1)(d-2p-4) \frac{\ee^{-a\Phi_0}}{V(T_0)}   \label{rrr} \\
L^2_0 &=& (d-p-3)(d-2p-4) \frac{\ee^{-a\Phi_0}}{V(T_0)}. \label{landlambda}
\eeqs

In the derivation we assumed $k\neq 0$. Also three 
special dimensionalities were excluded for compactness:
\begin{itemize}
\item[a)]
The case $d = p+3$ leads to an infinite radius in the AdS space-time,
i.e.~the flat Minkowski space times a circle, and is not considered in 
what follows. 
\item[b)]
For $d=2p+2$ the dilaton becomes a free parameter. Rather 
than its vacuum expectation value $\Phi_0$, the charge $k$ 
is determined from the equation of motion (\ref{dilaton})
\beqs
 k^2 = -2 V(T_0)~f(T_0)~ .
\eeqs
The rest of the formulae (\ref{tnaught}), (\ref{rrr})  and  (\ref{landlambda}) 
are still valid.
\item[c)]
We assumed that $V(T_0) \neq 0$. In addition to some completely Ricci flat 
solutions this condition also excludes the middle dimensional branes, 
for which we have $d=2p+4$. 
In these dimensions the radii are
\beqs
R_0^2 = L_0^2 = 4~(p+1)~ \frac{f(T_0)}{k^2}~,
\eeqs
Now $T_0$ is determined from $V(T_0) = 0$ (not $V^\prime(T_0)=0$)
and $\Phi_0$ from
\beqs
V'(T_0) \ee^{a\Phi_0} - \frac{k^2}{2} \frac{f'(T_0)}{f(T_0)^2} =0~.
\eeqs
\end{itemize}
The solutions discussed above are physically acceptable only for
$f(T_0)>0$.

\section{Dual field theory interpretation}

In this section we finally make contact with the conjectured gravity/field
theory duality by studying the field theory duals of the type 0 theories
for the simple case of $d=p+2$. This case already contains all the
relevant qualitative features of the most general one, without the 
complication of the Kaluza--Klein analysis. 

Let us state the logic of the approach. 
The classical solutions on the gravity side 
correspond to fixed points on the field theory side. 
There is a geodesic flow that relates these classical solutions, 
which is interpreted as the renormalization group 
flow connecting different fixed points. 
Fluctuation modes with positive, vanishing, and 
negative mass square correspond to irrelevant, 
marginal, and relevant deformations. 
The critical exponents can be obtained from these masses and they depend on
a finite set of undetermined parameters due to the arbitrariness of the 
tachyon couplings. These parameters should be fixed by comparing some 
universal quantities with experiment which leads to a prediction
for the remaining quantities.

\subsection{Stability of the AdS$_d$ solutions}

Classical solutions can serve as sound vacua for 
a quantum theory only if small fluctuations around the solutions 
are stable. In Minkowski space this implies that tachyonic 
fluctuation modes are forbidden. 
In an AdS background this requirement can be relaxed, and one
finds the bound 
\cite{Abbott, Breitenlohner, Mezincescu:a, Mezincescu:b}
\beqs
m^2 \geq - \frac{(d-1)^2}{4}~ \frac{1}{R_0^2}~. \label{BF}
\eeqs 
for the masses of the scalar fluctuation modes.

The first source of these instabilities near the solutions found in 
the previous section are obviously fluctuations 
in the tachyon field $T$. Tachyonic instabilities 
may enter also through the various scalar fields that appear on the AdS space, 
as the fields are compactified on the sphere $S^{d-p-2}$. 
In order to show that the theory is stable against these perturbations, 
one has to
linearize the full set of equations of motion around the classical 
solution, and check that no mode violates 
the bound (\ref{BF}).

This can be done, but the physically relevant features 
already appear in the case where the transverse sphere is absent. 
We shall discuss this example below in detail.

As $d=p+2$,  the nontrivial RR field is the top-form,
dual to a cosmological constant
\beqs
F_{\mu_1 \cdots \mu_{p+2}} &=& F~ \sqrt{-g}~ 
\epsilon_{\mu_1 \cdots \mu_{p+2}}~.
\eeqs
Note, that this is no longer the Freund--Rubin ansatz, but the RR 
field is a priori entirely general and unconstrained.
This is the field discussed in Section 3.3.
The equation of motion (\ref{formfield}) becomes 
in this case a constraint, and it turns out that 
the conserved charge is
\beqs
k = f(T)~\ee^{b\Phi}~F~. \label{cha}
\eeqs 
With the help of (\ref{cha}), the 
equations of motion for the other fields reduce to a Hamiltonian form
\beqs
\Box \Phi  &=& -\frac{\de}{\de\Phi} {\cal V}(\Phi,T) \\
\Box T &=& -\frac{\de}{\de T} {\cal V}(\Phi,T) \\
R_{\mu\nu}  &=& \frac{1}{2}\de_\mu\Phi\de_\nu\Phi+\frac{1}{2}\de_\mu 
T\de_\nu T-\frac{1}{d-2}{\cal V}(\Phi,T)g_{\mu\nu}
\eeqs
where the effective potential is 
\beq
{\cal V}(\Phi,T) = - V(T) \ee^{a\Phi} -
\frac{1}{2}~ \frac{k^2}{f(T)}~ \ee^{-b\Phi}~.
\eeq

Let us linearize the equations of motion near a classical solution
\beqs
\Phi &=& \Phi_0 + \varphi \\
T &=& T_0 + t \\
g_{\mu\nu} &=& \hat{g}_{\mu\nu} + h_{\mu\nu}~.
\eeqs
In order to do this, we need some knowledge of the functions 
$V(T)$ and $f(T)$. 
The only characteristics of these functions that will 
enter the stability analysis are the coefficients
\beqs
x = \frac{V'(T_0)}{V(T_0)}~, \qquad 
y = \frac{V''(T_0)}{V(T_0)}~, \qquad \mbox{and} \qquad
z = \frac{f''(T_0)}{f(T_0)} \label{xxyyzz}~.
\eeqs
Perturbative string theory analysis around $T=0$ yields \cite{Klebanov:a}
\beqs
V(T) &=& d-10 - \frac{d-2}{8}~T^2  + {\cal O}(T^4) \\
f(T) &=& 1 + T + \frac{1}{2}~ T^2 + {\cal O}(T^3).
\eeqs
This is not enough to determine the coefficients (\ref{xxyyzz}), 
and they should indeed 
be treated as free parameters of the theory. Including other unknown
functions (see discussion in Section 3.1) would give rise to more than three
such parameters, but the analysis performed here would still have the same 
qualitative features.

The fact that the graviton fluctuations actually decouple completely 
from those of the scalars simplifies the calculations: 
The graviton equations of motion can, in fact, be 
derived to first order from the effective action 
\beqs
S_{{\mathrm h}} &=& \int \dd^dx~ \sqrt{-(\hat{g}+h)}~ 
\Bigg\{ R(\hat{g}_{\mu\nu} + h_{\mu\nu}) + 
{\cal V}(\Phi_0,T_0) \Bigg\}~.
\eeqs
The scalar fluctuations obey
\beqs
\Big( -{\Box}+ 
{\cal M}~ \Big)
\left(\begin{tabular}{c} 
$\varphi$\\ 
$t$ 
\end{tabular}  \right)  =0 
\label{scalfluct}
\eeqs
where the mass matrix is
\beqs
{\cal M} = d(d-1)~ R^{-2}_0~ \left(\begin{tabular}{cc} 
$1$ & $\sqrt{\frac{d-2}{2}}\,x$ \\
$\sqrt{\frac{d-2}{2}}\,x$ & $d~ x^2 -y - \frac{2}{d} z$ 
\end{tabular}  \right)~.
\eeqs
The mass eigenvalues are
\beqs
m^2_{1,2} &=& d(d-1) \, R^{-2}_0 \, \Bigg(1 + \frac{\tau}{2} 
\pm \frac{1}{2}\sqrt{\tau^2+(2d-4) x^2}\Bigg)
\eeqs
where 
\beqs
\tau = d~ x^2 -\frac{2z}{d}-y-1~.
\eeqs
Note that the masses depend only on two independent parameters 
$x$ and $\tau$.

If we assume, following \cite{Klebanov:b}, that $f(T) = \exp(T)$, 
then the equations of motion give $x=-2/d$, 
and we can easily extract some interesting 
qualitative features as the only  
undetermined parameter is $\tau$. 

In this case 
there turns out to be three different, continuously connected  phases: 
First, there can be two 
particles, both with positive mass squared. Second, there can be a 
particle and a tachyon that obeys the bound (\ref{BF}). 
Third,  there can be a tachyon that makes the vacuum unstable. In AdS/CFT 
correspondence this translates into the statement that there can be 
at most one relevant operator in the infrared 
near the fixed point described by this theory.

\subsection{Solutions connecting conformal fixed points}

The stability analysis as applied to the critical points 
of the potential yields local information 
about the behavior of the dual field theory near its fixed points.
Depending on the form of the potential, there may exist gravity 
solutions that interpolate between different critical points.
These solutions should be interpreted on the field theory side as RG 
trajectories between conformal points.
 
In order to study these interpolating solutions we consider the 
ansatz   
\beq 
\label{Liouv-ansatz}
\dd s^2=\dd y^2+A^2(y)~ \dd x^2_{\parallel}
\eeq
and allow the two scalars to depend on the Liouville coordinate $y$.
We already know from the previous sections that there are exact 
solutions of the form
\beqs
A(y) = \ee^{y/R}~,
\eeqs 
where $R$ is the radius of the pertinent AdS space. 

The Einstein equation gives rise to two independent equations.
Defining  the following auxiliary function  
\beq    
\gamma(y)=(d-1)~ \frac{\dd}{\dd y}\log (A)~,
\eeq
the full set of equations takes the form 
\beqs
\frac{\ddot{A}}{A} + (d-2)\left( \frac{\dot{A}}{A} \right)^2 &=& 
\frac{1}{d-2}{\cal V} \label{dump1}\\
\ddot{\vec{{\Phi}}}+\gamma\dot{{\vec{\Phi}}} &=& 
-\vec{\nabla}{\cal V} \label{dump3} \\ 
\dot{\gamma} &=& 
-\frac{d-1}{2(d-2)}\big(\dot{\vec{\Phi}}\big)^2 \leq 0 \label{dump2}~.
\eeqs
Here we denote derivatives with respect to $y$ with a dot, 
and we have introduced the compact notation $\vec{\Phi}=(\Phi,T)$
for the two scalars.

Provided $\gamma \geq 0$, equation (\ref{dump3}) 
has the physical interpretation 
of a particle moving on a plane 
in the potential ${\cal V}$,  subject to a 
friction force. 
Let us assume that the potential has  two critical points 
$\vec{\Phi}_1$ and $\vec{\Phi}_2$ that satisfy 
${\cal V}(\vec{\Phi}_1) > {\cal V} (\vec{\Phi}_2)$, and that there is 
at least one unstable direction at $\vec{\Phi}_1$ for increasing $y$ 
and, similarly, a stable direction for $\vec{\Phi}_2$.
This can always be arranged by choosing the ${\cal O}(T^4)$ part 
in $V(T)$ suitably.
Due to the friction coefficient we expect our particle to roll down 
starting from the IR fixed point, and to  
converge in an infinite amount of time towards the lower UV fixed point.
This happens, since  $\gamma$ is strictly positive:
Indeed, at the critical points $\gamma$ 
approaches the values
\beqs
\gamma &\to& \frac{d-1}{R_1} \quad \mbox{for  } y\to -\infty \\
\gamma &\to& \frac{d-1}{R_2} \quad \mbox{for  } y\to +\infty~, 
\eeqs
and the friction coefficient decreases
monotonously between them according to (\ref{dump2}). 
This is consistent with the fact that
\beq
R^2_{1,2}=\frac{(d-1)(d-2)}{{\cal V}(\vec{\Phi}_{1,2})}
\eeq  
as follows from (\ref{dump1}).

The solution might be oscillatory near the UV critical point. 
Whether this happens depends on whether the friction  
is enough to stop the particle as it arrives 
at the lower point. 
Clearly, if one wants to interpret the result as an RG
flow, the oscillatory behavior would be difficult to accommodate 
in the field theory picture. The oscillatory solutions 
are exactly the solutions that would violate the bound (\ref{BF}), 
which is a necessary condition 
for the consistency of the system on the gravity side. 
Hence, quite remarkably, the stability in the gravity theory 
is dual to the consistency of the field theory interpretation.

Some universal information can be read from
the local behavior of these solutions. 
In the spirit of the Wilsonian RG treatment, let us study 
the critical behavior near the two fixed points 
in the linearized approximation. We must first identify the appropriate 
coordinate which in field theory can be consistently interpreted as 
the energy scale 
and parameterizes the interpolating solution. 
Such a coordinate can be chosen to 
be\footnote{This definition corresponds locally, near the fixed points, 
to the one  used in \cite{Maldacena}. However, 
there are alternative definitions. For instance choosing $U=\dot A$ 
one obtains the holographic relation, cf.~\cite{Peet}. All of these 
definitions lead to the same universal quantities.} 
\beq
U=\frac{A^2}{\dot{A}}
\eeq  
since at the critical points this reduces to $U=R~ \ee^{y/R}$, where 
the metric takes the standard form 
\beq
     \dd s^2 = \frac{R^2}{U^2} \dd U^2+\frac{U^2}{R^2} \dd x_{\parallel}^2~.
\eeq

Define
\beq
\vec{\Phi}(U) = \vec{\Phi}_0+\delta\vec{\Phi}(U)~,
\eeq
so that Eq.~(\ref{scalfluct}) takes the form
\beq
    \left[-\frac{1}{R^2}[d~U\de_U+U^2\de_U^2]+
    {\cal M}\right]\delta\vec{\Phi}=0.
\eeq
The eigenvalues of $\cal M$, namely  $m_i^2$ are 
found in Section 5.1, and for each of the two
eigenvectors we get two linearly independent solutions
\beq
\delta\tilde{\Phi}_i=A_iU^{\lambda^i_+}+B_iU^{\lambda^i_-}~,
\eeq
where 
\beq
\lambda^i_\pm=\frac{-(d-1)\pm\sqrt{(d-1)^2+4m_i^2R^2}}{2}~. \label{roots}
\eeq
Notice first that the stability condition (\ref{BF}) ensures 
the reality of the roots and they only depend on the dimensionless
parameters $x$, $y$, and $z$.

The IR limit corresponds to taking $U\to 0$, for which there must be at
least one positive eigenvalue, say, $(m^{{\mathrm IR}}_1)^2 >0$. 
In order for the solution not to 
blow up at this point we must choose $B^{\mathrm IR}_1=B^{\mathrm IR}_2=0$. 
If $(m^{{\mathrm IR}}_2)^2 <0$  we must 
also set $A^{\mathrm IR}_2=0$, otherwise, the trajectory may 
in general start with a
linear combination of the two eigenvectors. The trajectory will then evolve to
the UV fixed point as $U\to \infty$ where there will be at least one 
negative mass eigenvalue, say $(m^{{\mathrm UV}}_1)^2$. 
Generically, both the coefficients 
$A^{\mathrm UV}_1$ and $B^{\mathrm UV}_1$ will be non-zero and the root
$\lambda^1_+$ will dominate. 

In the AdS/CFT correspondence it is common to identify 
$g_{\mathrm YM}^2=\ee^\Phi$. This relation is plausible by considering
the weak coupling expansion of the world-volume gauge theory living on 
the stack of D-branes. On the other hand, away from the Gaussian fixed
point there is no a priori reason to make such an identification. Thus we
will stay general and regard $\Phi$ and $T$ as the coupling constants.

We can now read off the leading order 
behavior of the $\beta$-functions near the 
fixed points:
\beq
\beta_i(g_i)=U\frac{\dd g_i}{\dd U}=\lambda_+^i(g_i-g_i^*)+\dots,
\eeq
where $g_i=\tilde{\Phi}_i$. 
In particular, the conformal dimension of an operator 
coupled to 
the bulk field $\tilde\Phi_i(U,x)$ (a linear combination of the original 
tachyon and dilaton) is $\Delta_i= d-1 + \lambda_+^i$ and 
the anomalous dimension is $\lambda_+^i$. 

Note that, if $\lambda_+$ vanishes, then we should have $A=0$ as well.
This represents a marginal operator within our approximation and the 
computation of the $\beta$-function would pick up the sub-leading contribution
which in the UV is $\lambda_- = -(d-1)$. This is what happens 
in~\cite{Kehagias:a}, where the RG flow is studied in the framework of 
type II supergravity. 

One could also study confining and asymptotically free solutions of these 
models as well as extend to situations where Kaluza--Klein modes are 
present. We hope to return to some of these issues in the future.

\section{Acknowledgments}

We wish to thank R. Iengo, G. Mussardo, 
S. Randjbar-Daemi and A. Schwimmer for 
discussions. 
We are also grateful to all the participants to the
A. Salam I.C.T.P. journal club for providing a forum for discussion.
This work was supported in part by the European Union TMR programs CT960045
and CT960090.

\end{document}